\documentclass[]{spie}  %>>> use for US letter paper
\usepackage{amsmath,amsfonts,amssymb}
\usepackage{graphicx}
\usepackage{textcomp} % for \textmu, degree symbol, etc.
\usepackage{float} % for [H] figure placement
\usepackage[colorlinks=true, allcolors=blue]{hyperref}
\usepackage{wasysym}

\title{Linear motion flexure (``R-FLEX'') for miniature 6.2\,mm pitch optical fiber positioning robots with polar (R-\scalebox{1.5}{$\boldsymbol{\theta}$}) kinematics}

\author[a]{N.~R.~Wenner}
\author[a]{J.~H.~Silber}
\author[b]{M.~S.~Schubnell}
\author[a]{D.~J.~Schlegel}
\author[a]{R.~W.~Besuner}
\author[c]{W.~V.~Shourt}
\author[b]{A.~P.~Hope}
\affil[a]{Lawrence Berkeley National Laboratory, Berkeley, CA, USA}
\affil[b]{University of Michigan, Ann Arbor, MI, USA}
\affil[c]{University of California, Berkeley, CA, USA}

\authorinfo{Further author information: jhsilber@lbl.gov}

\begin{document}
\maketitle

\begin{abstract}
R-FLEX is a compact, low-part-count flexure-based radial positioning mechanism designed for the next generation of massively parallel fiber-fed spectroscopic telescope instruments. Current instruments such as the Dark Energy Spectroscopic Instrument (DESI) employ 5,000 robotic fiber positioners at 10.4\,mm pitch, whereas future surveys require 2.5--3$\times$ higher packing density, necessitating new positioning technologies. The R-FLEX mechanism converts small rotations at the flexure base into large, nearly tilt-free radial motion at the fiber tip, achieving naturally low-backlash linear motion within a compact \diameter5.8\,mm package. Coupled with a rotating $\theta$-stage, this enables overlapping circular patrol areas at 6.2\,mm pitch, as envisioned for projects like Spec-S5. Development proceeded through parametric modeling and optimization, finite element analysis, fabrication, and prototype testing. Prototype units characterized by optical centroiding achieved a corrected radial accuracy consistently better than 4\,\textmu m RMS over a $3.9$\,mm travel range, with maximum fiber tilt $0.092^\circ$, defocus 42\,\textmu m over the required range, and durability over more than 400,000 targets per robot across operating and survival temperature extremes. Preliminary Spec-S5 accuracy, defocus, and fiber tilt requirements are specified for the complete R-$\theta$ robot, and these radial-stage results exceed those requirements, leaving margin for the companion $\theta$-stage. A parametric optimization pipeline makes R-FLEX a versatile, mass-producible platform that can be re-optimized for precision-positioning applications beyond Spec-S5. This offers the precision, compactness, and reliability needed to collect hundreds of millions of spectra for new studies of the large-scale structure of the universe.
\end{abstract}


\section{Introduction}
\label{sec:intro}

\begin{figure}[H]
    \centering
    \includegraphics[width=\textwidth]{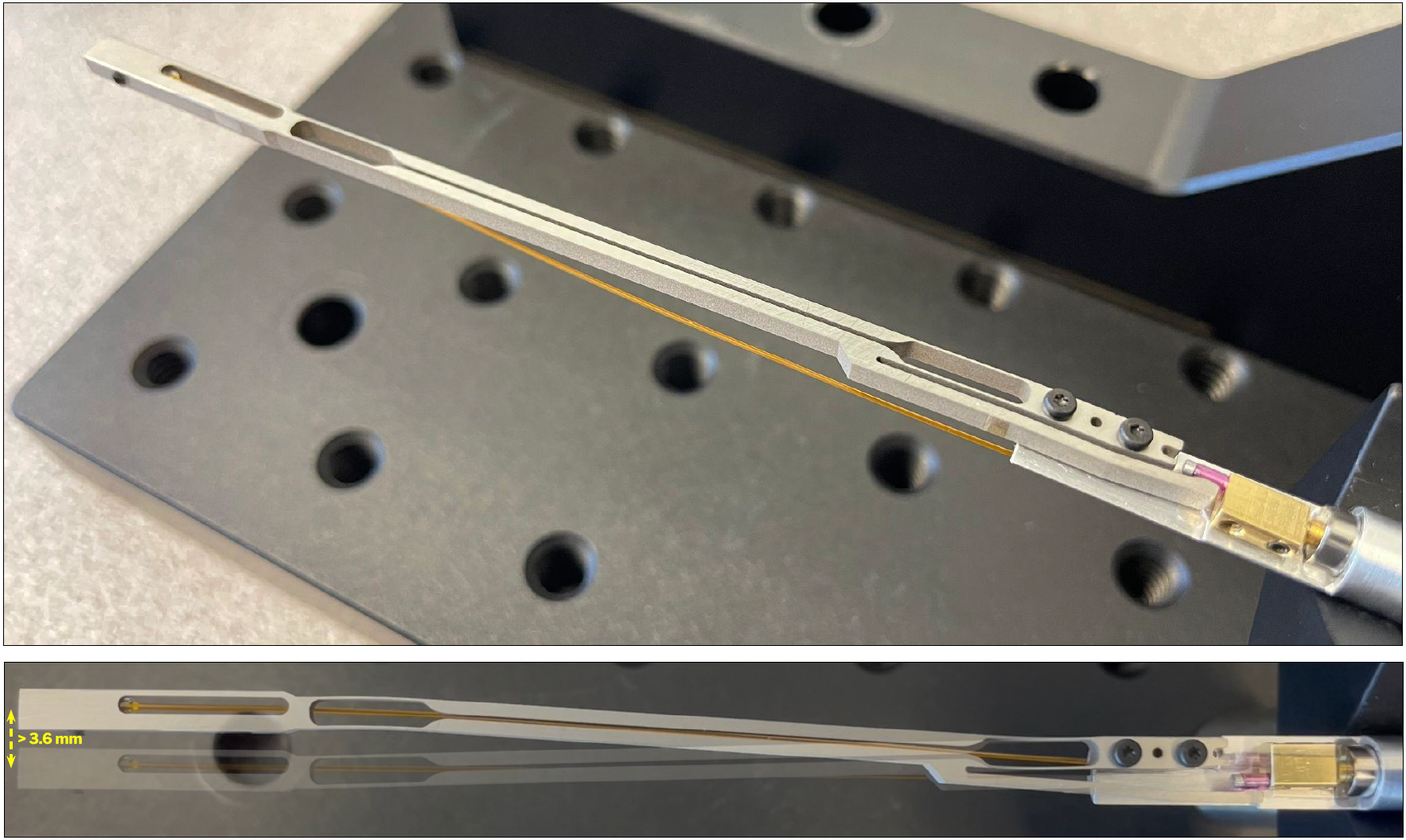}
    \caption{R-FLEX, with an overlay showing the fully retracted (transparent) and fully extended positions.}
    \label{fig:R-FLEX}
\end{figure}

\begin{figure}[H]
    \centering
    \includegraphics[width=0.75\textwidth]{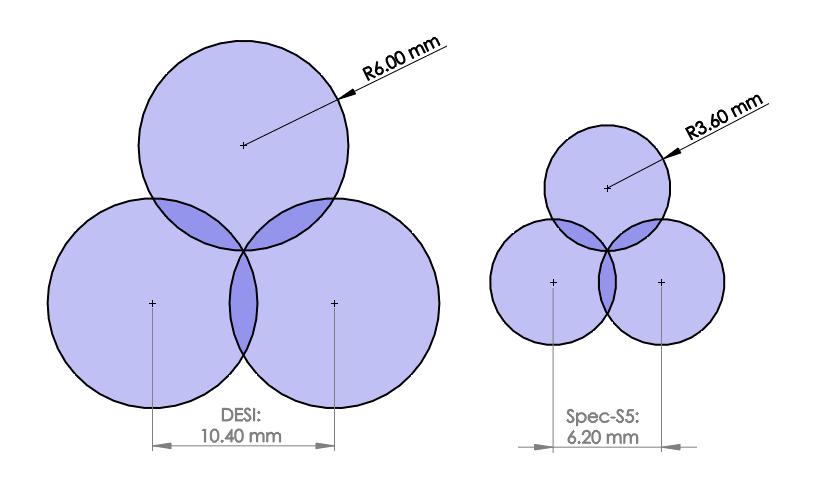}
    \caption{Next-generation spectroscopic surveys like Spec-S5 require 2.5--3$\times$ higher robot packing density compared to current surveys like DESI. Overlapping circular patrol areas allow for complete coverage of the focal surface.}
    \label{fig:patrols}
\end{figure}

The study of the large-scale structure of the universe is driven by ever-larger spectroscopic surveys. The Dark Energy Spectroscopic Instrument (DESI) uses 5,000 robotic fiber positioners packed at 10.4\,mm pitch and has collected over 56 million galaxy spectra to-date\cite{Silber2023}. Next-generation surveys such as Spec-S5 call for an order of magnitude more spectra, requiring 2.5--3$\times$ higher packing density and tens of thousands of positioners\cite{Silber2022, Schubnell2026} (Figure~\ref{fig:patrols}).

\begin{figure}[H]
    \centering
    \includegraphics[width=0.75\textwidth]{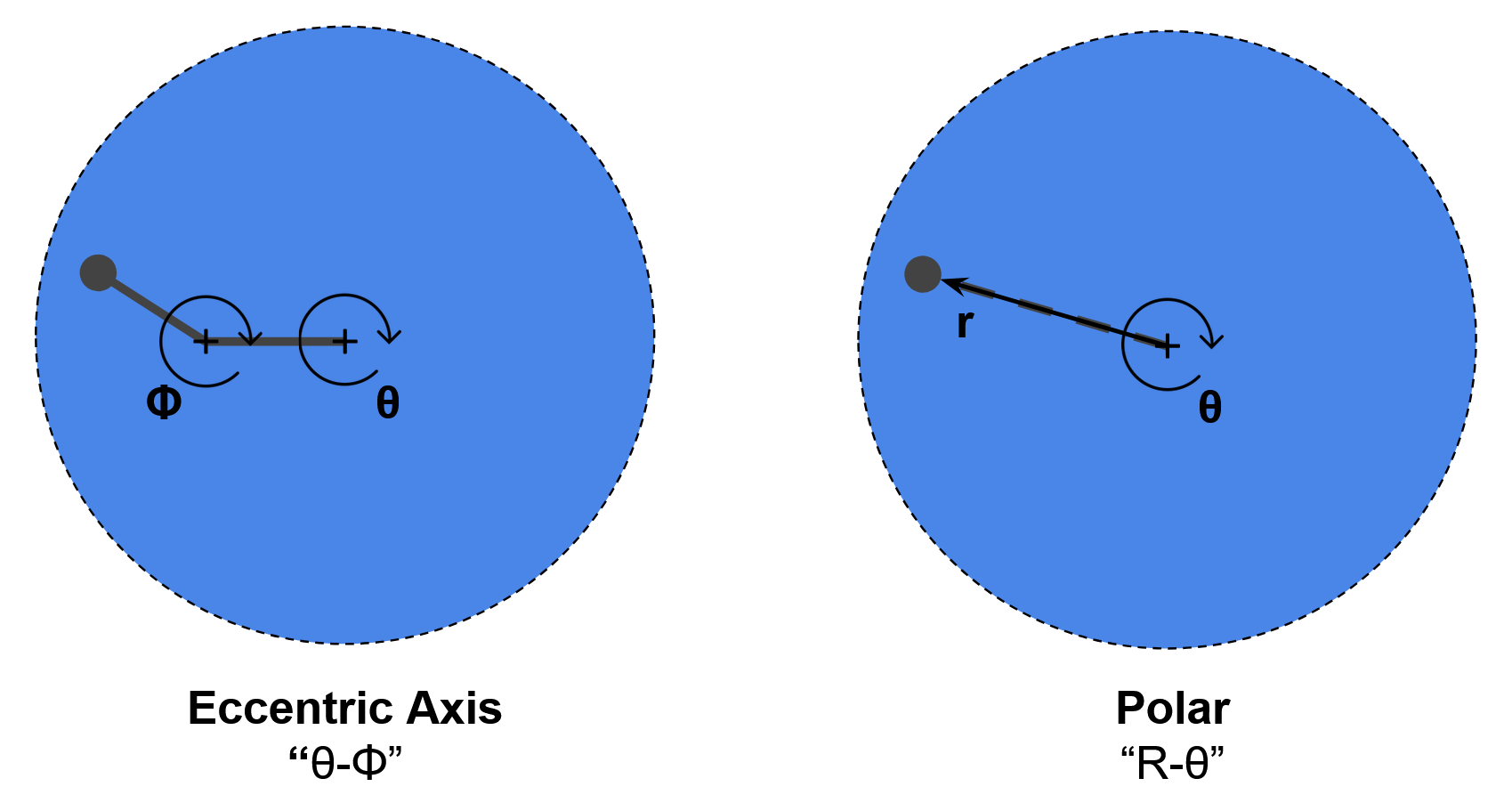}
    \caption{Two fiber-positioner kinematic schemes. R-FLEX adopts the polar ``R-$\theta$'' approach. Coupled with a rotating $\theta$-stage, R-FLEX enables full coverage of a circular patrol area.}
    \label{fig:kinematics}
\end{figure}

Many prior large-scale deployments such as those used for the DESI\cite{Leitner2018} (5,000x robots), LAMOST\cite{Cui2012} (4,000x), PFS\cite{Wang2024} (2,394x), and SDSS-V\cite{pogge_robotic_2020} (500x) instruments use a two-arm ($\theta$--$\phi$) kinematic scheme (Figure~\ref{fig:kinematics}) in which two rotary motors articulate a fiber across a circular patrol area. As one shrinks the pitch at which robots are mounted, the available cross-sectional area of the mechanical envelope shrinks as the square of the pitch. Mechanical packaging becomes quite challenging, even when using the current state-of-the-art in small, high-precision \diameter 4~mm brushless direct-current (BLDC) motors and gearheads. Gear backlash, part count, and per-unit cost become increasingly difficult to manage across the tens of thousands of units required. Alternately (as with PFS) piezo-based rotary motors have been utilized, with their own particular control and production challenges. Still another area of interest has been in tilting spine fiber robot technologies, such as those used in FMOS\cite{Akiyama2008} (400x). The spines can be packed very tightly, but also inherently tilt the fiber at significant and varying angles with respect to the chief ray across the patrol area.

The polar (R-$\theta$) kinematic scheme (Figure~\ref{fig:patrols}) has long been recognized as a compelling alternative\cite{Schlegel2008, Silber2012}. Calibration of as-built robots is simpler than with two rotational output axes, because the end effector motions are never degenerate for the two degrees of freedom. Anti-collision (planning motion paths to not collide with neighboring robots) is also simplified with R-$\theta$ motion because the common ``retract-rotate-extend'' strategy can be applied to a mechanism with truly linear retraction and extension (rather than its approximation by a rotating eccentric axis).

The present paper presents R-FLEX, a flexure-based radial positioning mechanism that adopts this polar (R-$\theta$) approach. R-FLEX converts small rotations at a flexure base into large, nearly tilt-free radial motion at the fiber tip, providing naturally low-backlash linear motion within a \diameter5.8\,mm envelope. The only backlash present is that of the gearmotor's gearbox driving the mechanism. Coupled with a rotating $\theta$-stage, R-FLEX produces overlapping circular patrol areas at 6.2\,mm pitch. We describe the mechanism, its parametric design and analysis, fabrication, and prototype performance, and show that R-FLEX prototype units meet all key Spec-S5 requirements while leaving margin for expected performance of a companion $\theta$-stage (the central rotation stage upon which an R-FLEX mechanism would be mounted). We do not discuss here any specific $\theta$-stage implementation.

\section{Mechanism}
\label{sec:mechanism}

\begin{figure}[H]
    \centering
    \includegraphics[width=\textwidth]{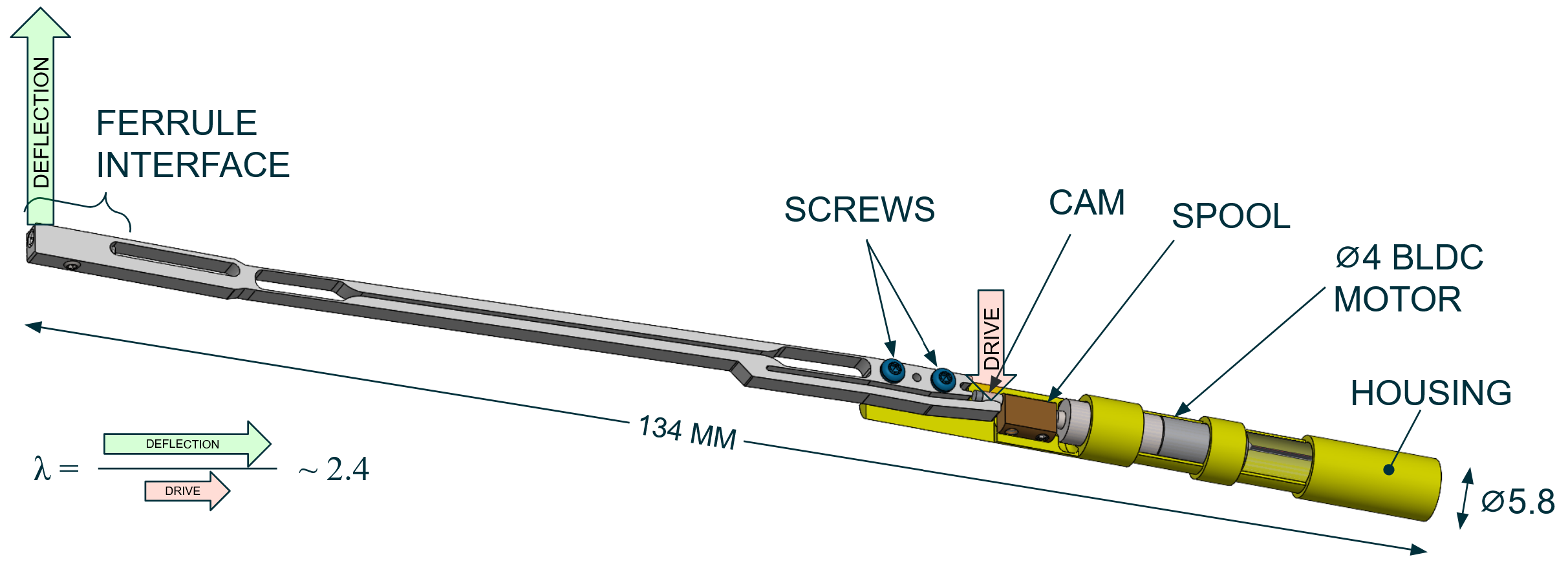}
    \caption{The R-FLEX assembly, showing the BLDC gearmotor, roller cam, lever arm, and four leaf flexures that together form a parallel linear spring. A lever arm amplifies the motion of the cam by the ratio $\lambda$.}
    \label{fig:assembly}
\end{figure}

R-FLEX converts small rotations at the flexure base into large radial motion with minimal tilt at the fiber tip. A roller cam, driven by a \diameter4\,mm BLDC gearmotor, pushes a lever arm, actuating four thin leaf flexures arranged as a parallel linear spring. This drives the fiber tip in the opposite direction with an amplification ratio $\lambda \approx 2.4$ (Figure~\ref{fig:assembly}). The result is nearly linear motion in space with natural backlash rejection and near-zero tilt.

\begin{figure}[H]
    \centering
    \includegraphics[width=0.8\textwidth]{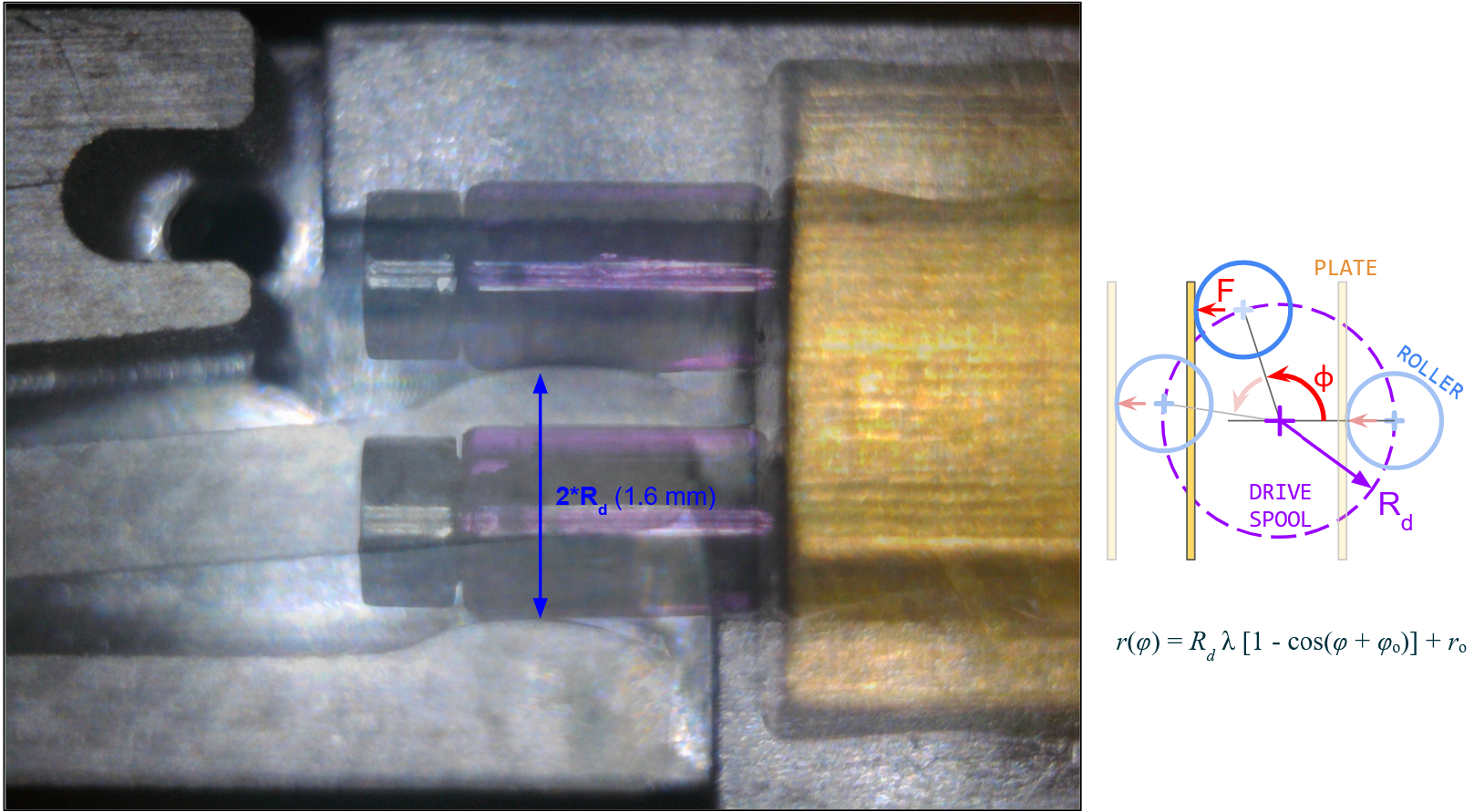}
    \caption{Radial fiber-tip position as a function of motor angle.}
    \label{fig:roller_kinematics}
\end{figure}

The radial fiber-tip position varies with cam angle as
\begin{equation}
    r(\varphi) = R_d\lambda\,\bigl[1 - \cos(\varphi + \varphi_0)\bigr] + r_0,
    \label{eq:kinematic_model}
\end{equation}
where $\varphi$ is the cam angle, $R_d$ is the radius at which the cam actuates the lever arm, $\lambda$ is the amplification ratio, $\varphi_0$ is an angular offset, and $r_0$ sets the radial origin (Figure~\ref{fig:roller_kinematics}). A cam rotation of $0$--$180^{\circ}$ produces the full range of the radial stroke at the fiber tip, with a maximum sensitivity at the fiber tip of $dr/d\varphi \approx 0.050$\,mm/deg near mid-travel. The gearmotor has a high reduction ratio (${\sim}337{:}1$ in the units currently used). The cam has a cylindrical profile, and is mounted eccentrically to the output shaft of the gearmotor. Our present prototypes use DESI drive electronics and  firmware, rotating in quantized open-loop steps with a minimum resolution of ${\sim}0.1^{\circ}$ at the motor shaft. This minimum step size corresponds to a maximum radial increment of ${\sim}0.015$\,\textmu m at the fiber tip---more than two orders of magnitude finer than the measured radial accuracy of $<$4\,\textmu m RMS (Table~\ref{tab:requirements}), such that step quantization is negligible. The gearmotor has no encoder, and radial position is commanded open-loop from a calibrated motor-angle-to-position map, which motivates a blind-move-plus-single-correction scheme described in Section~\ref{sec:methods}.

\begin{figure}[H]
    \centering
    \includegraphics[width=0.8\textwidth]{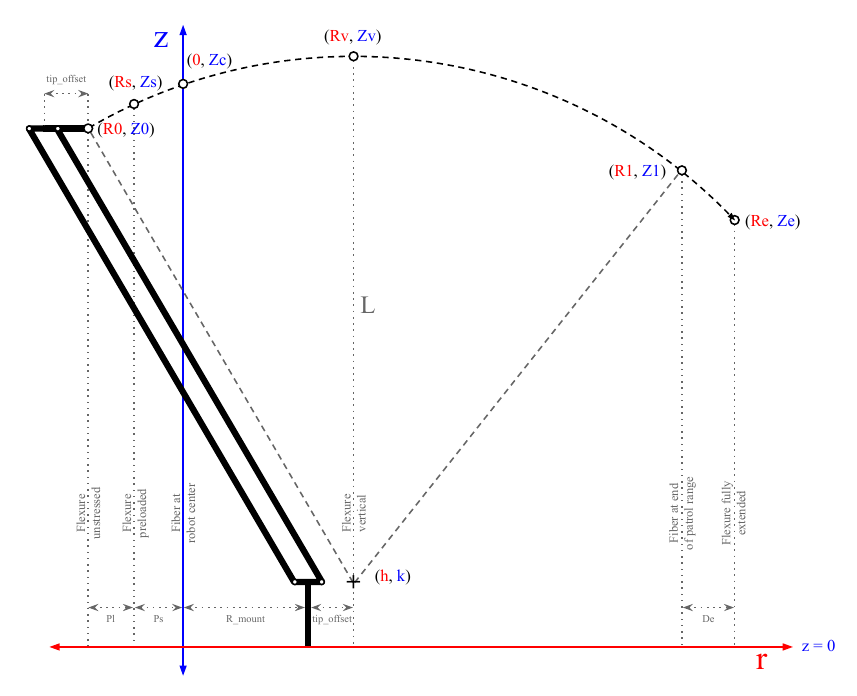}
    \caption{The flexure kinematics are similar to those of a four-bar parallel-motion linkage. The small out-of-plane excursion of the fiber tip over the travel range gives rise to defocus. The total defocus can be minimized by placing the vertical point of the flexure such that $R_\mathrm{v} = R_1 / 2$, although such placement may be constrained by available space.}
    \label{fig:defocus_kinematics}
\end{figure}

\section{Design and analysis}
\label{sec:design}

Candidate flexure materials were ranked by the ratio of elastic modulus to fatigue strength ($E/S_f$, where lower is better). Ti-6Al-4V was selected for its low $E/S_f$, low residual stress, and lack of heat-treatment requirements. The design was developed via a Python parametric sweep of $>$100,000 geometry variants filtered on stiffness, travel, defocus, stress, and manufacturability, then refined and validated with finite element analysis (FEA).

\begin{figure}[H]
    \centering
    \includegraphics[width=\textwidth]{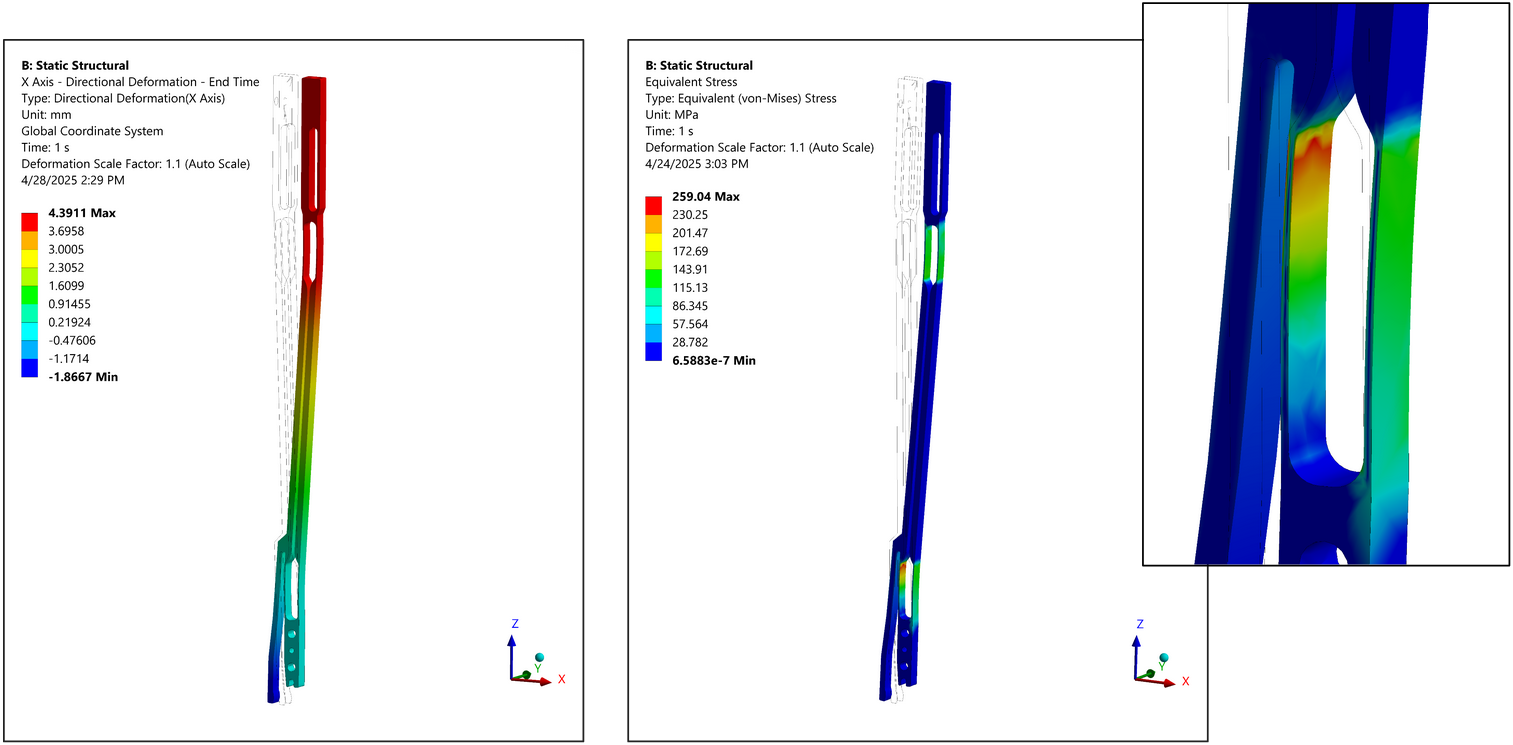}
    \caption{FEA of a typical load case at full travel, showing the von-Mises stress distribution with a peak of $\sim$260\,MPa concentrated in the leaf nearest the connection point with the lever arm.}
    \label{fig:fea}
\end{figure}

\begin{table}[H]
    \centering
    \caption{Factor of safety (FOS) results from FEA. The design lifetime requirement is $10^5$ cycles, with one target per cycle.}
    \label{tab:fea}
    \begin{tabular}{lc}
        \hline
        {\bf Quantity} & {\bf Value} \\
        \hline
        Max stress       & 260\,MPa \\
        Yield FOS        & 3.4 \\
        Fatigue FOS @ $10^5$ cycles & 2.5 \\
        Fatigue FOS @ $10^6$ cycles & 2.1 \\
        Fatigue FOS @ $10^7$ cycles & 1.9 \\
        \hline
    \end{tabular}
\end{table}

\section{Fabrication}
\label{sec:fabrication}

Flexures were produced by wire electrical discharge machining (wire EDM), which holds tight tolerances and can cut multiple parts in a single stacked operation. Remaining parts are computer numerical control (CNC) machined or commercial off-the-shelf (COTS). With 4 custom and 5 COTS parts, the design is suitable for mass production. Two complete prototypes (RFLX0001, RFLX0002) were built and tested.

\section{Performance}
\label{sec:performance}

Prototypes were characterized against the full set of Spec-S5 requirements, spanning radial accuracy, travel range, defocus, fiber tilt, lifetime, and operating and survival temperatures. Several of these requirements---radial accuracy, defocus, and fiber tilt---are specified for the complete R-$\theta$ robot, whereas the measurements reported here characterize the R-FLEX radial stage alone. For these quantities the radial stage consumes only part of the full-robot budget, leaving margin for the companion $\theta$-stage and for integration effects. The remaining requirements pertain directly to the radial stage. This section describes the test methods, presents the measured results, and discusses their implications. The requirements and corresponding radial-stage measurements are summarized in Table~\ref{tab:requirements}.

\subsection{Methods}
\label{sec:methods}

\begin{figure}[H]
    \centering
    \includegraphics[width=\textwidth]{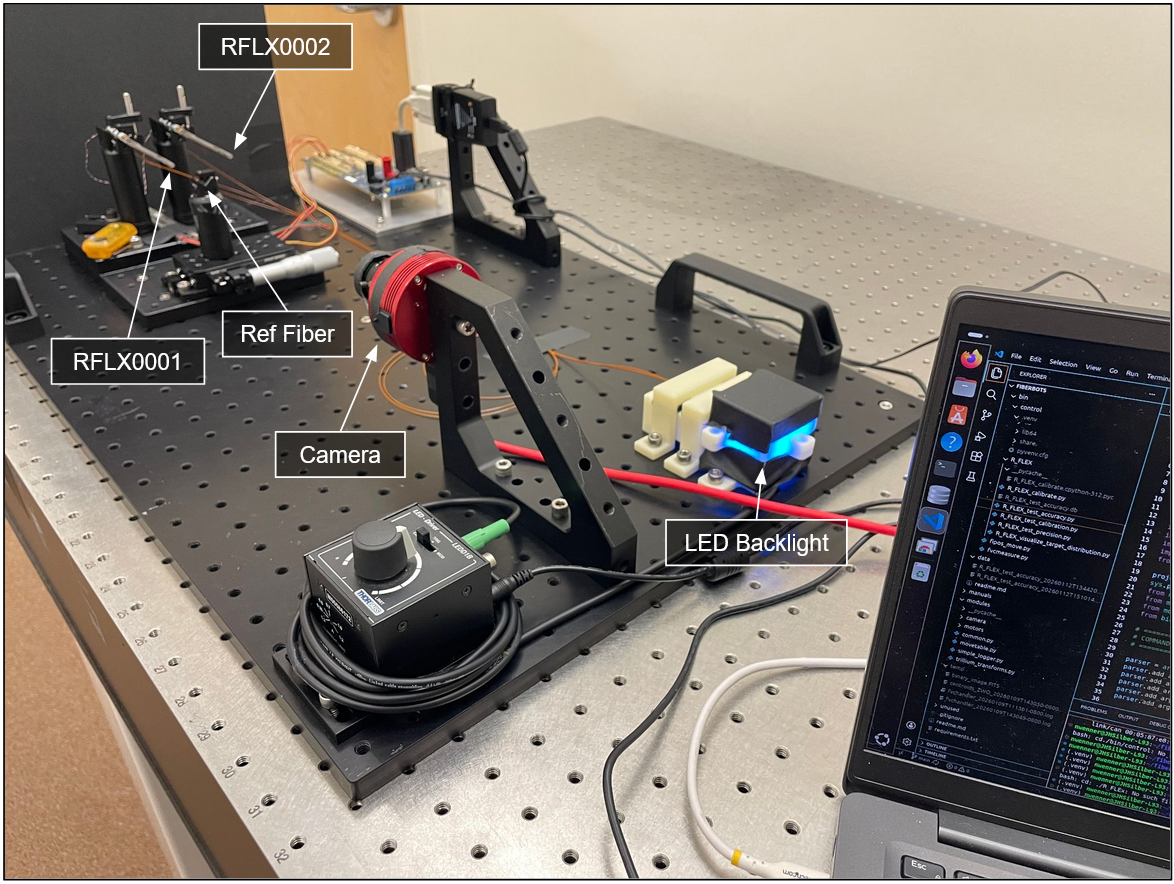}
    \caption{Test set-up for optical centroiding of RFLX0001 and RFLX0002 with back-lit fibers, used to measure radial accuracy over many thousands of targets.}
    \label{fig:set-up}
\end{figure}

The accuracy of both prototypes was characterized by optical centroiding of back-lit fibers (repeatability 0.3\,\textmu m root-mean-square, RMS) in tests ranging from 100--30,000 pseudo-random radial targets each. Targets were randomly drawn from a distribution that increases linearly with radial position. This simulates operating conditions with a uniform areal distribution of targets when paired with a rotating $\theta$-stage, where most targets occur at the outer (more extended, higher stress) range of the flexure.

At the beginning of each test, a dedicated calibration run established the motor-angle-to-fiber-position map for that robot. The three parameters of the kinematic model (Equation~\ref{eq:kinematic_model})---$R_d\lambda$ (treated as a single fitted parameter), $\varphi_0$, and $r_0$---were fit over a subset of the targets used for the accuracy measurement, and the radial ($r$) axis was defined as the best-fit line through the measured calibration points. The resulting per-robot map was then used to convert each target radial position into a commanded motor angle.

Each targeting operation comprised a first (``blind'') move with a deliberate undershoot (e.g. 15\,\textmu m) followed by a single, optically measured correction move. Because all targets are thereby approached from a single (extension) direction, gearbox backlash and other hysteresis effects are avoided. The reported accuracy reflects the performance of this move-and-correct procedure rather than the goodness of the calibration fit itself, which is reported separately (Figure~\ref{fig:residuals}). To assess the accuracy of blind moves alone, dedicated tests were performed with no undershoot and no corrections.

\begin{figure}[H]
    \centering
    \includegraphics[width=\textwidth]{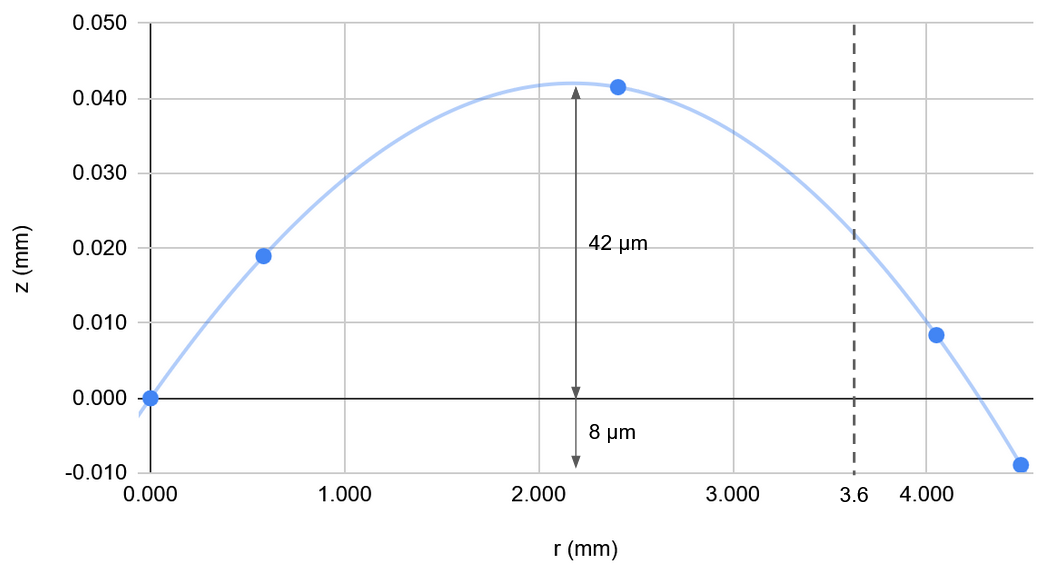}
    \caption{Measured defocus across the full travel range, reaching 42\,\textmu m at the 3.6\,mm required range---within the 50\,\textmu m requirement---and increasing to 50\,\textmu m at the end of the 4.2\,mm range.}
    \label{fig:defocus}
\end{figure}

Defocus (Figure~\ref{fig:defocus}) and angular tip and tilt were measured on an optical CMM at discrete points throughout the travel range of RFLX0001 only. This was done with the robot in a slightly different mechanical configuration than used for the accuracy measurements for that robot: the as-assembled radial position of the spool component varied slightly between set-ups, yielding effective travel ranges of approximately 4.2\,mm for the fiber tilt/defocus measurements and 3.9\,mm for the accuracy measurement. Both exceed the 3.6\,mm required travel range.

Both robots accumulated in excess of 400,000 targeting operations with at least 1 correction move over the course of development, distributed across the numerous test runs. These runs were conducted concurrently with iterative refinement of the control and analysis algorithms, and the results reported below were obtained after the algorithms had reached their refined state. Two runs are presented here as representative: a 1,000-target test (Figure~\ref{fig:convergence}) and a 10,000-target test (Figure~\ref{fig:accuracy}), conducted at cumulative counts of 399,743 and 387,643 operations at the beginning of each test, respectively---both well beyond the 100,000-target lifetime requirement.

Claude (Anthropic) was used to assist in developing the Python software for instrument control and data analysis. The authors specified the requirements, and all code was reviewed, tested, and validated by the authors.

\subsection{Results}
\label{sec:Results}

\begin{figure}[H]
    \centering
    \includegraphics[width=\textwidth]{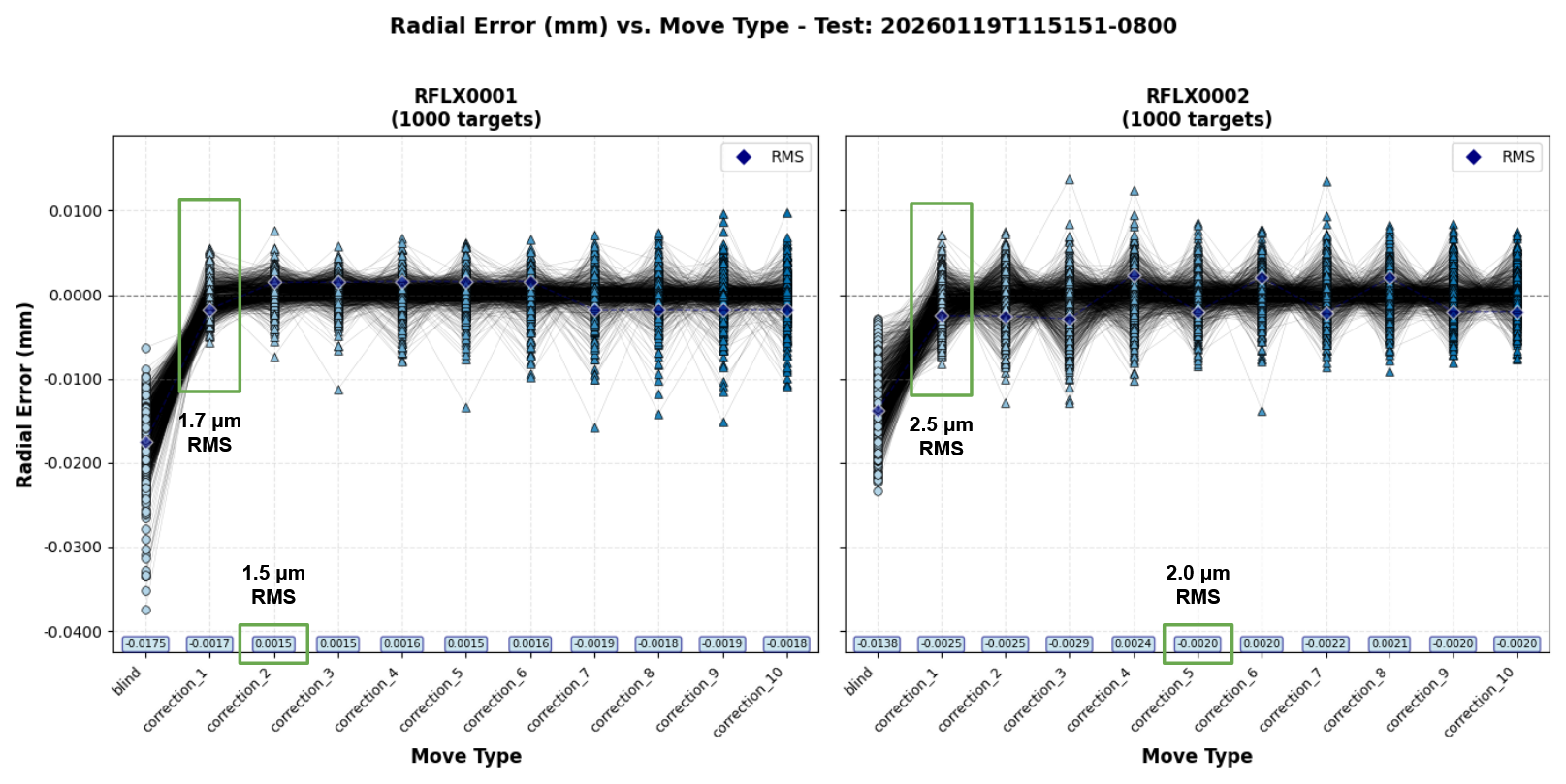}
    \caption{Representative radial accuracy performance. The first (``blind'') move intentionally undershoots, and the resulting position is measured optically then corrected. A single correction move is sufficient to meet the required 5\,\textmu m RMS accuracy, in this case achieving 1.7--2.5\,\textmu m RMS. With repeated corrections, peak accuracy ranged from 1.5--2.0\,\textmu m RMS after 2--5 corrections.}
    \label{fig:convergence}
\end{figure}

\begin{figure}[H]
    \centering
    \includegraphics[width=\textwidth]{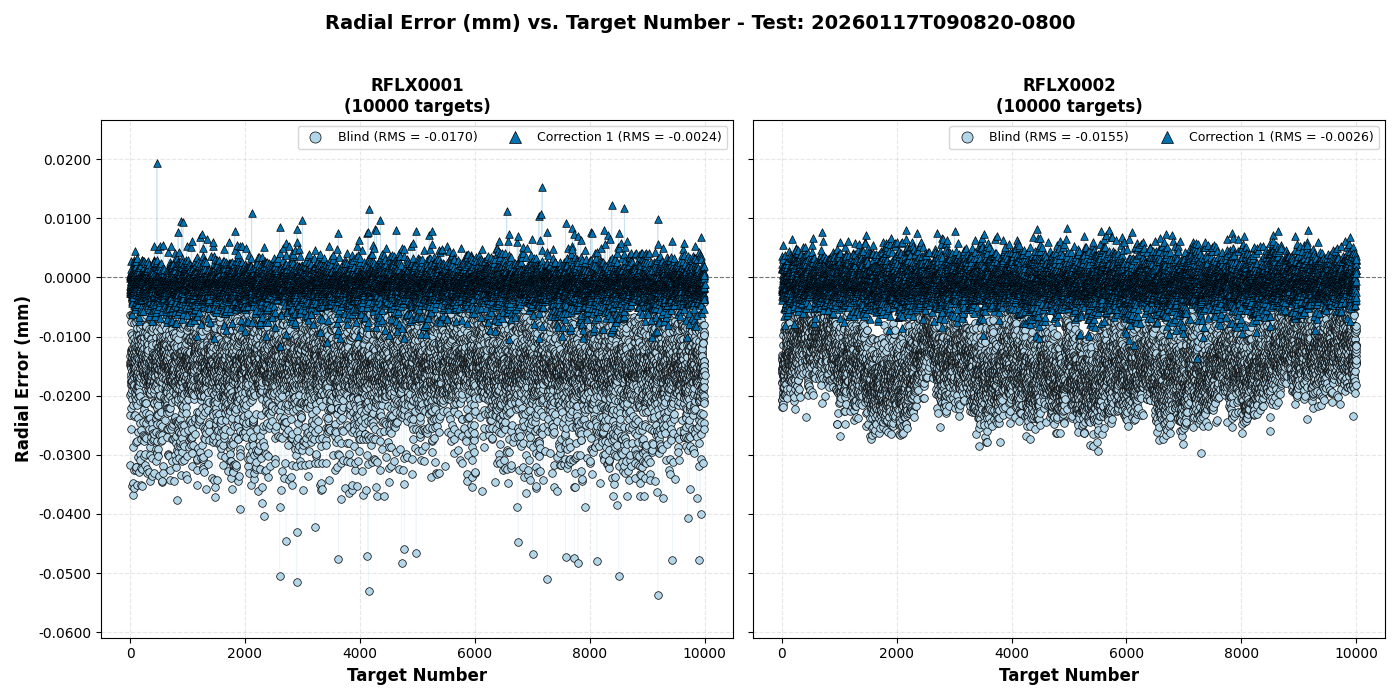}
    \caption{Representative radial accuracy after a single correction move. For 10,000 targets, the robots achieved radial accuracies of 2.4\,\textmu m RMS and 2.6\,\textmu m RMS. The full-robot requirement is 5\,\textmu m RMS. The blind move intentionally undershoots by 15\,\textmu m such that all correction moves approach their respective target from the same (extension) direction.}
    \label{fig:accuracy}
\end{figure}

Across the refined-algorithm tests, blind radial accuracy and corrected radial accuracy were consistently better than 6\,\textmu m RMS and 4\,\textmu m RMS, respectively---both well within the 50\,\textmu m RMS and 5\,\textmu m RMS respective requirements. The two representative tests achieved corrected radial accuracy of 1.7--2.5\,\textmu m RMS (Figure~\ref{fig:convergence}) and 2.4--2.6\,\textmu m RMS (Figure~\ref{fig:accuracy}).

\begin{figure}[H]
    \centering
    \includegraphics[width=\textwidth]{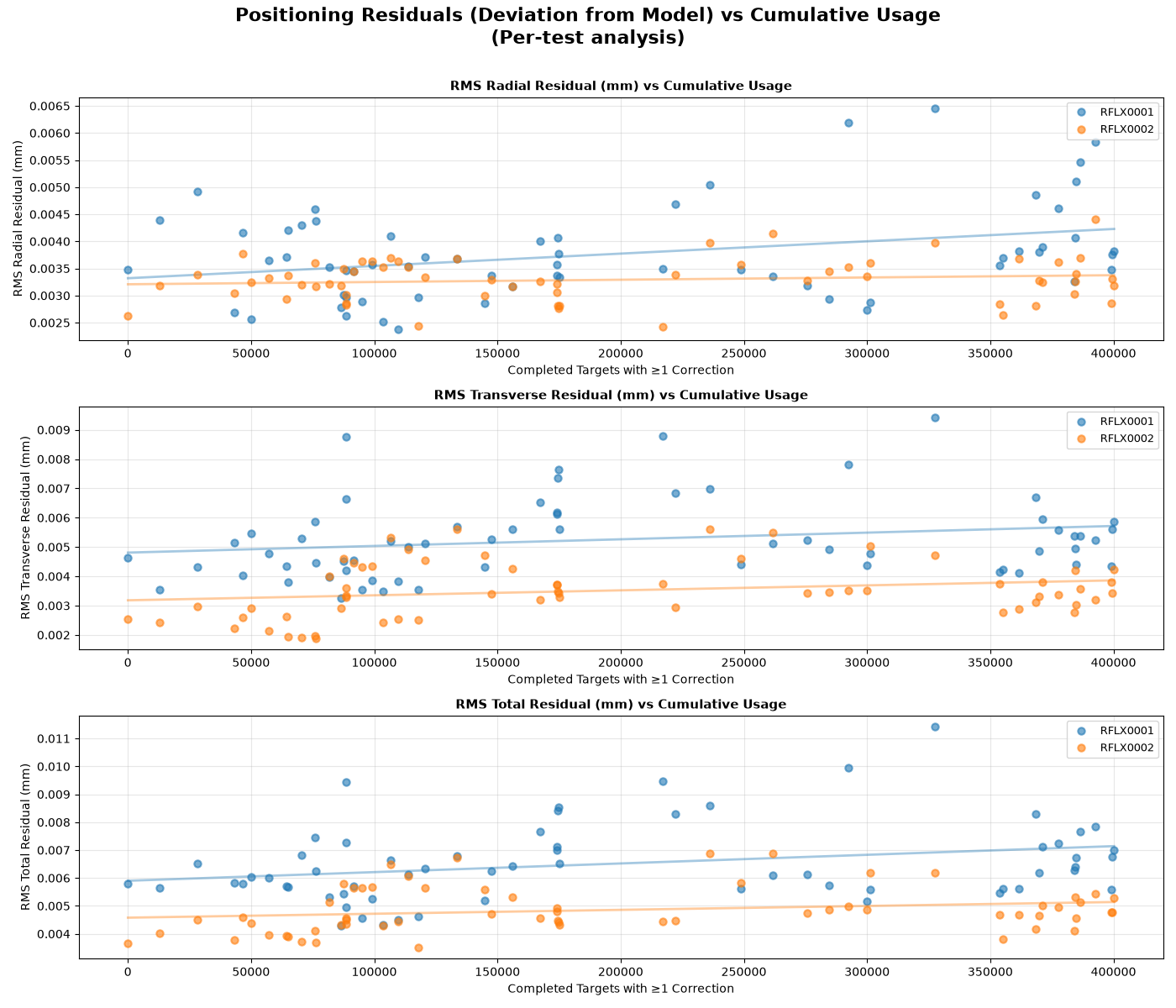}
    \caption{Positioning residuals (deviation from the calibration model) versus cumulative usage, with one point per test. The radial ($r$) axis is defined by the best-fit line through the calibration points for a given test.}
    \label{fig:residuals}
\end{figure}

Deviations from the positions predicted by the calibration model are shown in Figure~\ref{fig:residuals}, with each point summarizing one test. Observed fiber paths were linear (radial) to within 10\,\textmu m RMS. Radial residuals were less than 7\,\textmu m RMS for all tests. Note that the achieved radial accuracy is smaller than might be assumed from the radial residuals, since systematic errors from the calibration model are largely removed by a well-calibrated correction move. Over cumulative usage, the radial residuals trended weakly upward for RFLX0001 (a rise of less than 2\,\textmu m RMS across $\sim$400,000 targets), while RFLX0002 was essentially flat (rise $<$1\,\textmu m RMS).

RFLX0001 was measured to have 42\,\textmu m defocus and a maximum fiber tilt of 0.092$^{\circ}$ across the required range against full-robot allocations of 50\,\textmu m and $0.5^{\circ}$, respectively (Table~\ref{tab:requirements}). Maximum fiber tilt was calculated as the quadrature sum of the maximum measured orthogonal components: tip (0.02$^{\circ}$) and tilt (0.09$^{\circ}$).

Performance proved durable and robust, with radial accuracy sustained over $>$400,000 targets per robot and maintained from $-21\,^{\circ}$C to $+41\,^{\circ}$C operating temperature and through $-40\,^{\circ}$C to $+70\,^{\circ}$C survival cycling.

\begin{table}[H]
    \centering
    \caption{Preliminary requirements for Spec-S5 fiber robots are given alongside corresponding values for the measured performance of the R-FLEX radial stage. In rows marked~$^{\dagger}$, the requirement applies to the complete R-$\theta$ robot. In each of these, the radial stage consumes a portion of the full-robot budget, leaving margin for the $\theta$-stage as well as integration tolerances. The fiber-tilt value is the quadrature sum of the maximum measured orthogonal tip and tilt components. Defocus and fiber tilt were measured for one prototype unit, whereas all other values were measured for two prototype units.}
    \label{tab:requirements}
    \begin{tabular}{lll}
        \hline
        & {\bf Required} & {\bf Measured} \\
        \hline
        Envelope          & $\leq$ \diameter5.8\,mm                & \diameter5.8\,mm \\
        Pitch             & 6.2\,mm                                & 6.2\,mm \\
        Radial range      & $\geq$ 3.6\,mm                         & 3.9\,mm \\
        Blind accuracy$^{\dagger}$    & $\leq$ 50\,\textmu m RMS                  & $<$ 6\,\textmu m RMS \\
        Corrected accuracy$^{\dagger}$& $\leq$ 5\,\textmu m RMS       & $<$ 4\,\textmu m RMS \\
        Defocus$^{\dagger}$           & $\leq$ 50\,\textmu m       & 42\,\textmu m \\
        Fiber tilt$^{\dagger}$        & $\leq 0.50^{\circ}$            & 0.092$^{\circ}$ \\
        Lifetime          & $\geq$ 100,000                         & $>$ 400,000 \\
        Operating temp    & $-20\,^{\circ}$C to $+40\,^{\circ}$C   & At least: $-21\,^{\circ}$C to $+41\,^{\circ}$C \\
        Survival temp     & $-30\,^{\circ}$C to $+60\,^{\circ}$C   & At least: $-40\,^{\circ}$C to $+70\,^{\circ}$C \\
        \hline
    \end{tabular}
\end{table}

\subsection{Discussion}
\label{sec:Discussion}

\paragraph{Fatigue margin.}
The fatigue factors of safety in Table~\ref{tab:fea} are likely conservative for two reasons. First, the analysis adopts a conservative S--N curve for Ti-6Al-4V. Second, it assumes that every move reaches peak stress, whereas in operation moves traverse only a portion of the full travel range and therefore induce lower stress (although the realized stress distribution is expected to skew toward the high end). Taken together, these factors suggest that the mechanism retains substantial fatigue margin beyond what Table~\ref{tab:fea} indicates. This is consistent with the experimental result that radial accuracy was sustained over $>$400,000 targets per robot---several times the 100,000-target lifetime requirement. This margin informs the second-generation design described in Section~\ref{sec:status}.

\paragraph{Accuracy and the correction move.}
Blind-move accuracy was consistently below 6\,\textmu m RMS---nearly an order of magnitude better than the 50\,\textmu m requirement and approaching the 5\,\textmu m requirement for corrected moves. A single correction move was sufficient to meet requirements and reduced the error to consistently below 4\,\textmu m RMS (Figure~\ref{fig:convergence}, Figure~\ref{fig:accuracy}). Additional corrections can yield further improvement, with peak accuracy of 1.5--2.0\,\textmu m RMS after 2--5 corrections in one test (Figure~\ref{fig:convergence}). That one correction move suffices is significant for survey operations: the number of correction iterations directly sets focal-plane reconfiguration time, and limiting it to one preserves observing efficiency at scale.

In practice, the operational accuracy of a newly deployed unit is expected to be at least as good as reported here. The reported accuracy was obtained from late-life runs (near 400,000 cumulative targets) under a refined correction scheme, whereas comparably tuned measurements are not available for early-life runs. The residuals of Figure~\ref{fig:residuals} are computed by applying a single, consistent calibration method to every test, so they track the intrinsic repeatability of the mechanism independently of the correction scheme that was live at test time. On this basis, the residuals of RFLX0001 drifted weakly upward over cumulative usage while those of RFLX0002 remained essentially flat, indicating that the mechanism was at worst marginally less repeatable over time, and for one unit essentially unchanged. Because corrected radial accuracy is limited in large part by mechanism repeatability, a newer unit would be expected to perform comparably or slightly better.

\paragraph{Linearity, fiber tilt, and defocus.}
Observed fiber paths were linear to within 10\,\textmu m RMS, confirming that the four-leaf parallel-flexure arrangement closely approximates ideal radial motion. In practice, these residuals will be corrected by the companion $\theta$-stage.

The radial stage's maximum fiber tilt of 0.092$^{\circ}$ contributes less than $20\%$ of the full-robot $0.5^{\circ}$ budget.

Among the quantities specified at the full-robot level, defocus carries the least margin: 42\,\textmu m against a $\leq$50\,\textmu m requirement. Two points are worth noting: The ideal flexure vertical-point placement that minimizes defocus ($R_\mathrm{v} = R_1/2$, Figure~\ref{fig:defocus_kinematics}) is not achievable within the available envelope, so the present result is close to the practical minimum for this geometry. Second, the stated budget is a preliminary requirement adopted from DESI experience, and the final Spec-S5 allocation may be more or less stringent. The companion $\theta$-stage must fit within whatever margin remains, and defocus may therefore drive future design trades.

\begin{figure}[H]
    \centering
    \includegraphics[width=\textwidth]{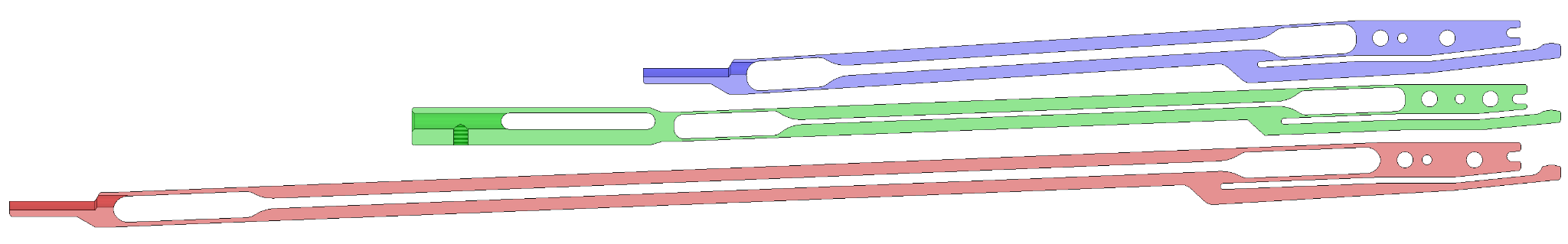}
    \caption{The R-FLEX flexure (green) and a new design (blue) optimized for stiffness and reduced length. Also shown is a configuration optimized for longer ($>$7.2\,mm) travel range (red).}
    \label{fig:designs}
\end{figure}

\paragraph{Adaptability.}
Although the present design targets Spec-S5, the flexure is highly adaptable. The same parametric optimization pipeline used to derive the baseline geometry can be re-run against different stiffness, travel, and defocus targets, making R-FLEX a candidate for a range of precision-positioning applications. For example, as shown in Figure~\ref{fig:designs}, the flexure can be shortened to optimize for stiffness or lengthened to enlarge the patrol area, and a variant optimized for longer ($>$7.2\,mm) travel is feasible while maintaining the same slim profile.

\paragraph{Limitations and open questions.}
The present results characterize two prototypes (RFLX0001, RFLX0002) rather than a production population, so unit-to-unit variation and yield statistics remain to be established across a larger build. Full validation as an integrated robot---R-FLEX combined with the companion $\theta$-stage---is still in progress (Section~\ref{sec:status}), and so the radial performance reported here does not yet include any contribution from $\theta$-axis pointing. Because the accuracy, defocus, and fiber tilt requirements are full-robot allocations, the integrated robot must combine the radial-stage contributions reported here with those of the $\theta$-stage. To remain within requirements, radial accuracy may need to be improved depending on the achievable accuracy from the $\theta$-stage. One open question is the extent to which transverse and radial accuracy are affected by frictional dynamics (e.g. stick-slip), and therefore improvable by reducing friction at key interfaces such as that between the roller and its shaft.

\section{Status and next steps}
\label{sec:status}

A companion $\theta$-stage (``Iris'') has been built and tested. Second-generation R-FLEX and Iris designs are complete and currently being fabricated, and they will be combined and tested as a full robot assembly. Upgrades to the second-generation R-FLEX robot include improved surface finish and reduced friction at the interface of the roller and its shaft. The second-generation R-FLEX flexure was re-optimized to be stiffer and shorter, deliberately operating at a lower fatigue factor of safety to exploit the conservative margin identified in Section~\ref{sec:Discussion}.

\section{Conclusions}
\label{sec:conclusions}

R-FLEX prototype units meet all key requirements for Spec-S5 in a compact, mass-producible package and sustain that performance over $>$400,000 targets and across temperature extremes. For accuracy, defocus, and fiber tilt requirements, which are specified at the full-robot level, the radial stage reserves margin for the companion $\theta$-stage. Its slim, low-part-count, low-backlash design is a strong candidate for mass deployment in next-generation multi-object spectroscopic surveys, and its parametric design pipeline makes it readily adaptable to other precision-positioning applications.

\acknowledgments

This work was supported in part by the Director, Office of Science, Office of High Energy Physics of the US Department of Energy under contract No.\ DE-AC02-05CH11231, and by the Laboratory Directed Research and Development (LDRD) program of Lawrence Berkeley National Laboratory. The authors acknowledge the contributions of the broader Spec-S5 collaboration and the many institutions supporting the development of next-generation spectroscopic survey facilities. This work builds upon the heritage of SDSS, BOSS, eBOSS, DES, DESI, and related survey projects.

The authors used Claude (Anthropic) to assist with language editing and manuscript organization. All content was reviewed and verified by the authors, who take full responsibility for it.

\end{document}